\documentclass[]{aa}
\usepackage{graphics}

\def\ie{{\it i.e.\/}\ }
\def\c31{${\it C_{31}}$}

\def\re{$r_e$}

\begin{document}


\title{1.65~$\mathrm{\mu m}$ (H-band) surface photometry of galaxies. IX:
photometric and structural properties of galaxies.}


\author{M. Scodeggio\inst{1}
\and G. Gavazzi\inst{2}
\and P. Franzetti \inst{1,2}
\and A. Boselli\inst{3}
\and S. Zibetti \inst{2}
\and D. Pierini\inst{4}  
}

\offprints{M. Scodeggio}
\mail{marcos@ifctr.mi.cnr.it}

\institute{Istituto di Fisica Cosmica ``G. Occhialini'', CNR, 
via Bassini 15, 20133, Milano, Italy
\and
Universit\`a degli Studi di Milano - Bicocca, P.zza 
dell'Ateneo Nuovo 1, 20126 Milano, Italy
\and 
Laboratoire d'Astrophysique de Marseille, Traverse du Siphon, F-13376 
Marseille Cedex 12, France
\and
Dept. of Physics and Astronomy, University of Toledo,
2801 W. Bancroft, 43606, Toledo, Ohio, USA
}

\date{Received 31 July 2001 / Accepted 8 January 2002}

\titlerunning{Photometric and structural properties of galaxies}
\authorrunning{M. Scodeggio et al.}

\maketitle

\begin{abstract}
As a result of a systematic NIR $H$-band ($1.65\mu$m) imaging
survey of normal galaxies in the local universe that includes objects
both in the Virgo cluster and in the ``Great Wall'' (including A1367,
A1656 as well as the ``isolated'' population in the bridge between the
two clusters), we are able to measure in a highly homogeneous way
photometric and structural properties for a sample of 1143
galaxies. We base our analysis on a quantitative structural parameter,
the concentration index C$_{31}$ (defined as the ratio between the
radii that enclose 75\% and 25\% of the total luminosity), instead of
relying on the galaxies' morphological classification. The C$_{31}$
parameter provides a model independent, quantitative and continuous
characterization of the light distribution within galaxies, and it is
thus to be preferred to either the Hubble type or a parameter like the
bulge-to-disk or bulge-to-total light ratio.  Low C$_{31}$ objects are
typically found among disk galaxies, while high C$_{31}$ describes
bulge-dominated systems.  We confirm our previous claim that C$_{31}$
correlates strongly and non-linearly with the the galaxy total
luminosity.  C$_{31} > 5$ values are found only at
L$_H>10^{10}$L$\odot$ (giant ellipticals mixed with early-type
spirals), while at L$_H<10^{10}$L$\odot$ galaxies have C$_{31}< 3$
(dwarf Irregulars mixed with ellipticals).  At high luminosity, low
C$_{31}$ are allowed (bulge-less giant Scs).  Thus C$_{31}$ and the
total luminosity are not sufficient to fully characterize the family
of galaxies.  However we find that galaxies can be completely
described by three parameters, namely: a scale parameter (the H-band
luminosity), a shape parameter (the concentration index C$_{31}$) and
a colour parameter (e.g. the B-H colour). At low luminosity dEs and
dIs, having similar C$_{31}$, are colour-discriminated, while at very
high luminosity different C$_{31}$ discriminate S0s from Scs,
otherwise undistinguishable on the basis of their colour.  A single,
monotonic relation exists between luminosity and $mu_e$ in the H-band,
as opposed to the two separate regimes that are generally observed in
the B-band.  As NIR luminosity traces quite accurately the galaxy mass
distribution, this relation re-enforces the indication in favour of a
scale-dependent mass collapse mechanism which produces higher
surface-brightness and more centrally peaked galaxies with increasing
mass.  However, the presence of high-luminosity low-C$_{31}$ galaxies
hints at other machanisms and physical properties (such as angular
momentum) playing an important role in galaxy formation.

\end{abstract}

\section{Introduction}
\label{sec:intro}

The most commonly used galaxy classification system is the one
originally developed by Hubble (\cite{realm}), and brought to
completion by Sandage (\cite{atlas}). Among the main reasons for its
success are the fact that it is relatively easy to use, and the fact
that it has proven successful in ordering the ``realm of nebulae''
according to important physical characteristics. The correlations
between quantities related to a galaxy stellar population (like galaxy
color) and Hubble type are well known (see Roberts \& Haynes
\cite{RH94} for a recent review).  However, the Hubble classification
system suffers from a number of important drawbacks that are becoming more
and more evident as the samples of galaxies studied in detail
become larger both at low and high redshift. Among these drawbacks, its
qualitative and quantized nature, that prevents fitting relations as a
function of type on a solid statistical basis, and the fact that a
large (and ever-growing, with the completion of new surveys) fraction
of galaxies cannot fit well inside the scheme, are the most limiting
features.

These and other limitations of the Hubble system have been recognized
for a long time, and a number of extensions to the scheme, or
alternative classification schemes, have been proposed through the
years. Best-known among such efforts are the extension to the Hubble
system devised by de Vaucouleurs (\cite{deVa59}), and the alternative
scheme proposed by Morgan (\cite{Mor58}, \cite{Mor59}). Still, neither
of these schemes is widely used by the astronomical community. More
recently, Whitmore (\cite{Whit84}) proposed a sound, quantitative
scheme for the classification of spiral galaxies, that is based on two
parameters: ``scale'' and ``form'' (following Whitmore's
terminology). The scale parameter is a combination of the galaxy
absolute blue luminosity and isophotal radius, while the form is a
combination of the galaxy color and bulge-to-total light ratio. This
idea has been further developed by Bershady et al. (\cite{BJC};
hereafter BJC), who have presented a classification scheme
encompassing all morphological types. In doing so, they were forced to
increase the dimensionality of the classification volume from two to
three dimensions, that they define to be a spectral index (galaxy
color), a scale (galaxy surface brightness), and a form (galaxy light
concentration, or asymmetry) parameter.  This classification scheme is
fully quantitative, can be used to classify galaxies over a broad
range of redshifts (provided that images with enough spatial
resolution can be obtained), and is quite successful in separating
normal galaxies into classes that are largely coincident with those
one would obtain using Hubble types.

One important feature for a classification scheme should be its
ability to describe not only the ``current'' status of a galaxy (by
``current'' we mean the status at the epoch of the observation, the
latter being of course dependent on the galaxy redshift), but also its
evolutionary history. This goal however is complicated by the
degeneracy that many galaxy parameters show with respect to different
possible formation and evolution histories. A well-known example is
the age-metallicity degeneracy for stellar populations in early-type
galaxies (Worthey \cite{Wor94}), but we observe a rather similar
color-star formation history degeneracy for late-type, star-forming
galaxies (Gavazzi et al. \cite{Gav01b}). To achieve the goal of
obtaining a classification scheme that might provide information on a
galaxy's past evolutionary history we are building a large dataset of
multifrequency observations of nearby galaxies, that covers objects of
all morphological types over an extended range of luminosities and
environmental conditions, from the cores of dense clusters to
relatively under-dense regions in the field. We began our effort by
focusing on late-type star-forming galaxies, and the results of the
analysis of these systems are summarized in Gavazzi et
al. (\cite{Gav96c}), and Gavazzi \& Scodeggio (\cite{GS96}). Since
then we have extended our sample to also cover early-type galaxies,
and we have completed a photometric near-infrared ($H$-band) imaging
survey that covers a large fraction of our total sample, to measure in
a completely homogeneous fashion the photometric and structural
properties of nearby normal galaxies. Previous papers in this series
describe the photometric survey (Gavazzi et al. \cite{Gav96a} -- Paper
I; Gavazzi et al. \cite{Gav96b} -- Paper II; Boselli et
al. \cite{Bos97}; Gavazzi et al. \cite{Gav00a} -- Paper III; Boselli
et al. \cite{Bos00} -- Paper IV; Gavazzi et al. \cite{Gav01a} -- Paper
VII) and provide details on the derivation of the photometric
parameters used here (Gavazzi et al. \cite{Gav00b} -- Paper V). In
this work we describe the observed correlations between photometric
and structural properties for the galaxies in our sample. For the
first time this kind of analysis is presented using $H$-band
near-infrared imaging data. This offers two key advantages with
respect to the more often used optical bands imaging: a much smaller
influence from dust extinction on the observed parameters, and the
capability of tracing the bulk of the luminous matter mass, without
being heavily influenced by the small fraction of young stars that
often dominate galaxy emission in the blue optical bands. Two other
papers in this series discuss the correlation between their
photometric, structural, and dynamical properties (Pierini et
al. \cite{Dan01}), and the correlation between these properties and
the galaxy star formation history (Boselli et al. \cite{Bos01}). A
future paper will discuss the practical application of our
classification scheme to a sample of more distant galaxies.
\bigskip

\section {The data}
\label{sec:data}

The analysis of photometric and structural properties of normal
galaxies we carry out in this paper is based on a nearly complete Near
Infrared $H$-band survey of an optically selected sample of nearby
galaxies. Full details about the sample selection, the observations,
and the data reduction procedures can be found in the previous papers
of this series (Papers I, II, III, IV, VII), and here only a brief
summary is presented.

\begin{figure*}
\resizebox{\hsize}{!}{\includegraphics{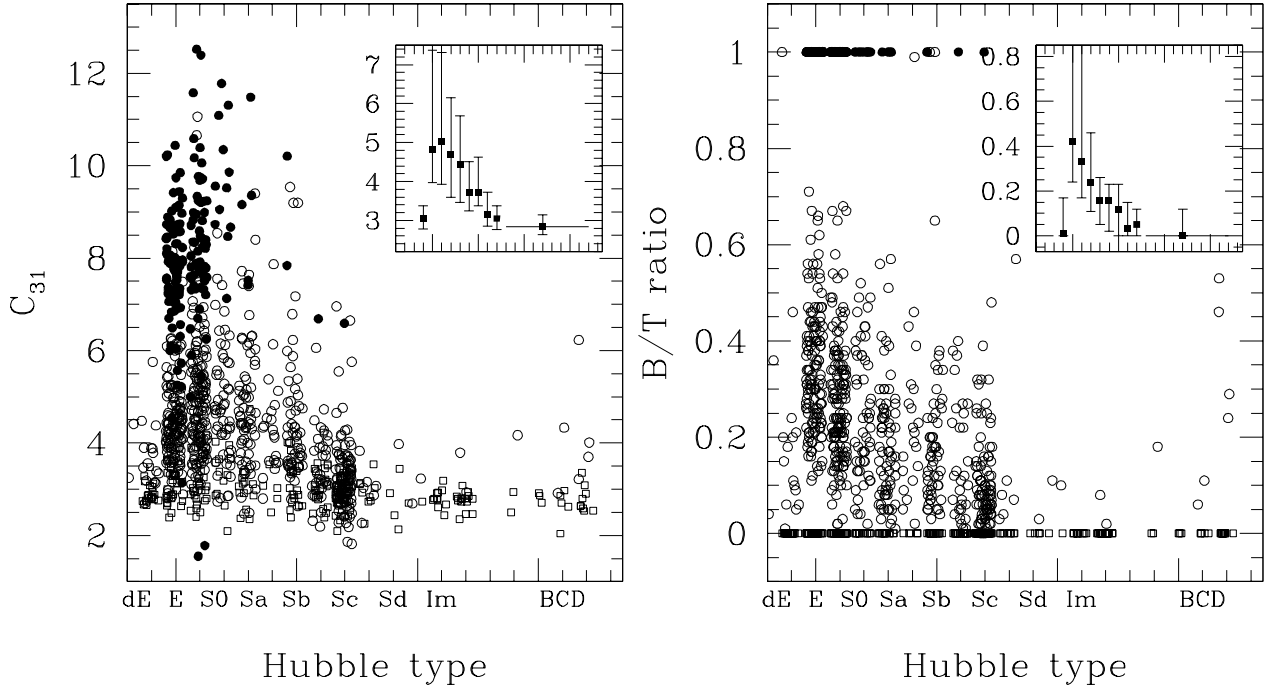}}
\caption{The distribution of concentration index C$_{31}$ values (left
panel), and of bulge to total light ratio B/T values (right panel), as
a function of Hubble type.  Filled circles represent galaxies with a
pure de Vaucouleurs radial profile, empty squares galaxies with a pure
exponential disk profile, and empty circles galaxies with a composite
profile. To make the density of data points more readily visible, we
offset the abscissa of each point in the graph by a small, random
amount.  The insets show the median C$_{31}$ and B/T value for each type
(filled squares), and the upper and lower quartiles of their
distributions.  Types later than Sc are grouped together to improve
statistics.}
\label{fig:c31_bt_type}
\end{figure*}

Our sample includes all galaxies from the CGCG catalogue (Zwicky et
al. \cite{CGCG}) (m$_\mathrm{pg} \le$~15.7) that belong to the Coma
supercluster region, including the two clusters Coma (A1656) and
A1367, and to the clusters A262 and Cancer, and galaxies from the VCC
catalogue (Binggeli et al. \cite{Bing85}) (restricted to m$_\mathrm{pg}
\le$~16.0) in the Virgo cluster. Observations were carried out from
1993 to 1997 with the 1.5~m TIRGO and with the Calar Alto 2.2 and
3.5~m telescopes equipped with the NICMOS3 $256^2$ pixel arrays
cameras ARNICA and MAGIC, respectively, and more recently with the ESO
NTT and TNG telescopes, equipped with the SOFI and NICS or ARNICA
cameras. In total we have obtained observations for 1302
galaxies, that span all morphological types from E and dE to
dIrr. However, because of the selection criterion of our master
catalog, extremely low surface brightness galaxies are not included in
our sample. Paper VII provides information on the completeness of our
observations.

The surface photometry measurements used in this paper are discussed
in detail in Paper V. Each galaxy two-dimensional light distribution
was fitted with elliptical isophotes to derive a radial surface
brightness profile. This profile was fitted with either a de
Vaucouleurs $r^{1/4}$ or an exponential disk profile, or a combination
of two profiles (either a de Vaucouleurs plus exponential disk, or two
exponential profiles, the choice being based on a best fit
criterion). Among the 1302 profiles, 192 are fitted with a pure de
Vaucouleurs $r^{1/4}$ law and 369 with a pure exponential one. The
remaining profiles require a Bulge+Disk (B+D) decomposition, with 54
objects showing signs of a truncated outer light distribution.  We
have also tried to fit the global surface brightness profile with a
single S{\'e}rsic (\cite{Sersic}) $r^{1/n}$ profile, but we could not
reproduce at all the shape of the observed profile for a large
fraction of the B+D galaxies, and therefore abandoned this experiment.
A total magnitude $H_\mathrm{T}$ was derived for each galaxy by
extrapolating the suitable profile to infinity, and adding the
extrapolated light contribution to the one measured within the last
fitted isophote. Empirical effective radius $r_\mathrm{e}$ (the radius
within which is contained half of the galaxy light) and effective
surface brightness $\mu_\mathrm{e}$ (the mean surface brightness
within the effective radius) were measured on the observed profile, at
the half light point, on the basis of the total magnitude
determination. Bulge and disk parameters were derived from the fitted
profiles, together with the bulge to total (B/T) light flux (the ratio
of the bulge flux to the total flux from the galaxy). We also derived
for each galaxy an isophotal radius $r_\mathrm{H} (20.5)$ determined
in the elliptical azimuthally--integrated profiles as the radius at
which the surface brightness reaches 20.5 H--mag~arcsec$^{-2}$, and a
concentration index (C$_{31}$), defined as in de Vaucouleurs
(\cite{deVa77}) to be the model--independent ratio between the radii
that enclose 75\% and 25\% of the total light $H_\mathrm{T}$. The
median uncertainty in the determination of the total magnitude is 0.15
mag, while those on the determination of log~$r_\mathrm{e}$ and
$\mu_\mathrm{e}$ are 0.05 and 0.16 mag, respectively.

\begin{figure}
\resizebox{\hsize}{!}{\includegraphics{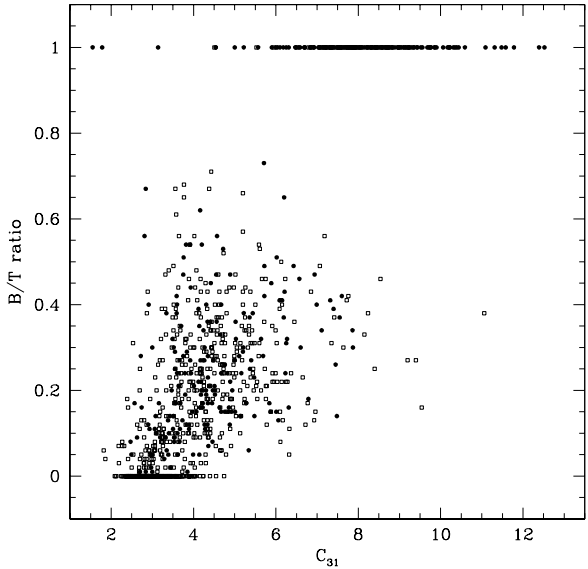}}
\caption{The relation between the concentration index C$_{31}$ and the
bulge-to-total light ratio for the galaxies in our sample. Same symbols
as in the previous figures.}
\label{fig:bt_c31}
\end{figure}

Although we do not use Hubble types as a basis for the classification
of galaxies, we have tried to leave the possibility of a comparison
with the traditional classification scheme open. For this reason in
the following analysis we exclude from our sample a number of galaxies
that have very uncertain morphological classification (90), and
peculiar galaxies (69). The total sample used here is therefore
composed of 1143 galaxies. Of these, 831 have color measurements
available, and are used in the definition of our classification scheme
(see Sect.~\ref{sec:cube}).

\section{The concentration index}
\label{sec:c31}

A concentration index has been used before as a classification tool,
most noticeably in the Morgan classification scheme (Morgan
\cite{Mor58}, \cite{Mor59}). Other classification schemes involving a
concentration index have been proposed, usually trying to take
advantage of the correlation existing between image concentration and
mean surface brightness in galaxies ({\it e.g.} Okamura et
al. \cite{Oka84}; Kent \cite{Kent85}; Doi et al. \cite{Doi93}; Abraham
et al. \cite{Abr94}). Most recently BJC have presented a global
classification scheme for galaxies that includes a concentration index
as one of the fundamental classification parameters, along with color,
surface brightness, and galaxy asymmetry. All their parameters were
derived from $B$-band imaging data. They find correlations to be
present between concentration and either color, surface brightness, or
asymmetry, the strongest one being that between concentration and
color.

Here we adopt a definition of concentration index which is somewhat
different from the one used by BJC. We define, following de
Vaucouleurs (\cite{deVa77}), the concentration index C$_{31}$ as the
ratio between the radii that enclose 75\% and 25\% of the total galaxy
light $H_T$, whereas BJC take the logarithm of the ratio between two
radii (specifically, those that enclose 80\% and 20\% of the total
light). As we discuss in section~\ref{sec:cube}, we consider our
``linear ratio'' definition better suited to emphasize the changes of
fundamental galaxy properties as a function of galaxy luminosity, as
discussed already in Gavazzi et al. (\cite{Gav96c}). With our
definition of concentration index, a pure de Vaucouleurs profile
should have C$_{31}$ = 7.02, while a pure exponential disk profile
should have C$_{31}$ = 2.80.

It is quite natural to expect a correlation between the value of
C$_{31}$ and other parameters traditionally used for galaxy
classification, like Hubble type, or bulge-to-disk / bulge-to-total
light ratios, as the relevance of the bulge inside a galaxy is one of
the main parameters in the Hubble scheme, and the only reliable one for
edge-on spiral galaxies. We show in the left panel of
Fig.~\ref{fig:c31_bt_type} the distribution of C$_{31}$ values as a
function of Hubble type. Two characteristics are clearly evident: there
is a general correlation between C$_{31}$ and type, as evidenced by the
global decline in the median C$_{31}$ value going from ellipticals to
late type spirals (with dE galaxies representing an exception with
respect to this trend; see the Figure inset), but there is also a large
scatter within a given type, at least up to type Sb. Part of this
scatter might result from misclassifications, due to the greater
difficulty in classifying galaxies at the distance of the Coma
supercluster with respect to those in the Virgo cluster, but a component
of this scatter is due to a real dependence of C$_{31}$ on galaxy
luminosity, as discussed in section~\ref{sec:cube}. It is therefore
clear that concentration index and Hubble type are not completely
equivalent classification criteria. 

\begin{figure*}
\resizebox{\hsize}{!}{\includegraphics{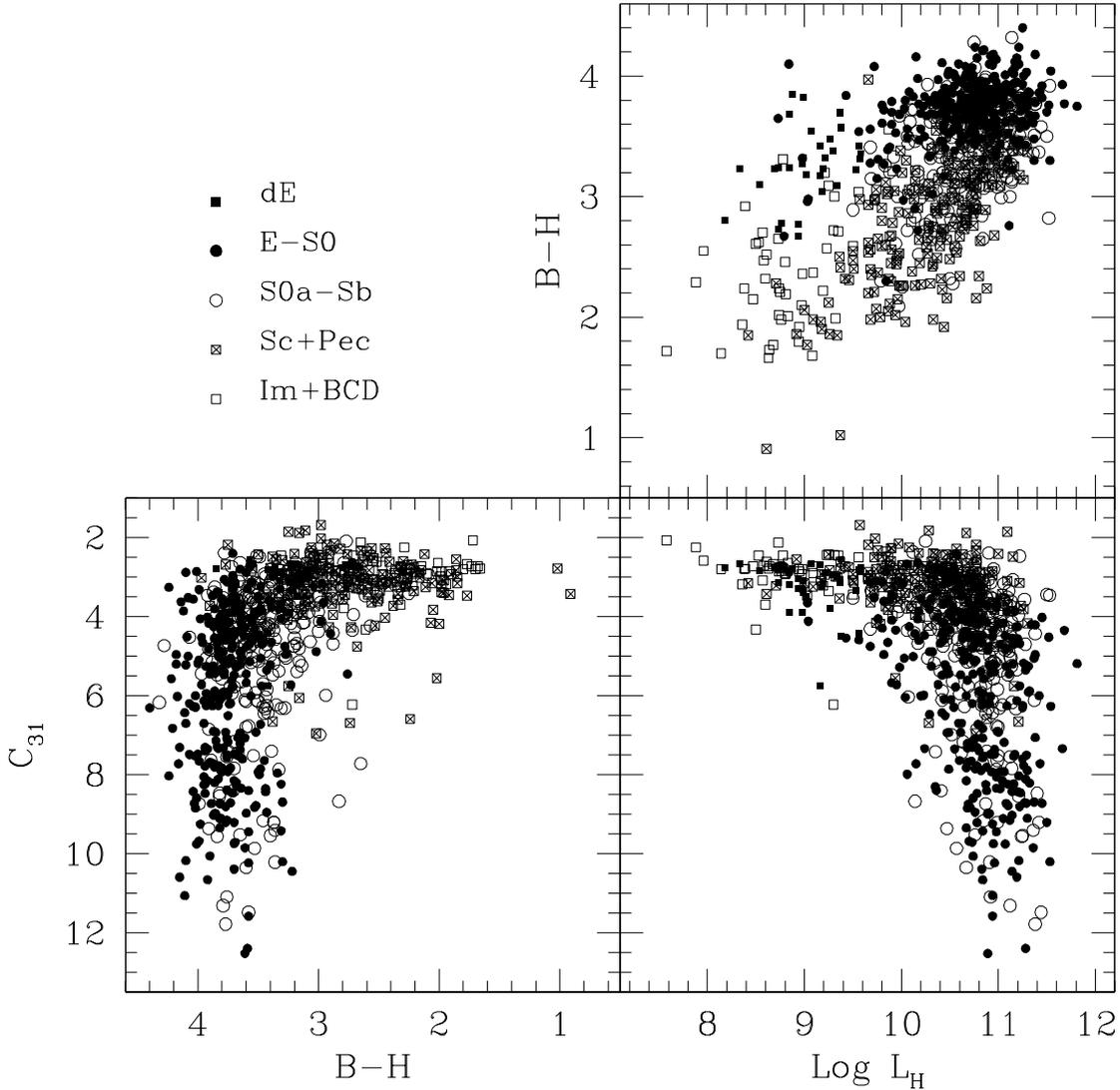}}
\caption{The distribution of $H$-band luminosity $L_H$, $B-H$ color,
and concentration index C$_{31}$ for the galaxies in our
sample. Different symbols identify galaxies within different ranges of
Hubble types, as detailed in the legend.}
\label{fig:lcube}
\end{figure*}

\begin{figure*}
\resizebox{\hsize}{!}{\includegraphics{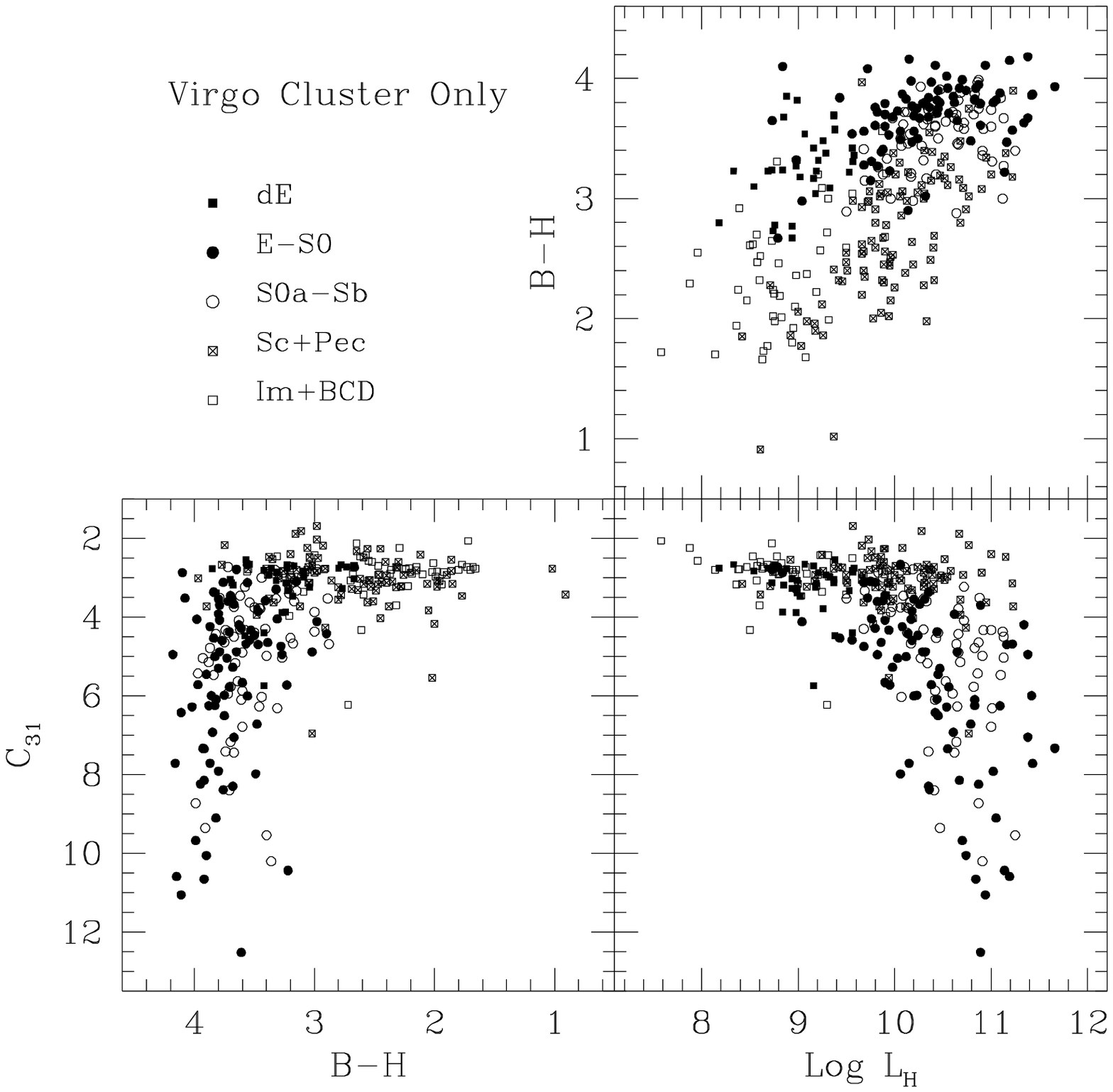}}
\caption{The distribution of $H$-band luminosity $L_H$, $B-H$ color,
and concentration index C$_{31}$ as in the previous figure, but
limited to Virgo Cluster galaxies only.}
\label{fig:lcube_virgo}
\end{figure*}

The same situation is present when considering the B/T light ratio. As
shown in the right panel of Fig.~\ref{fig:c31_bt_type} there exists
also for B/T a global correlation with Hubble type, with a decline in
the median value of B/T when going from ellipticals to late type
spirals, and a large scatter within each Hubble type. Also in this
case there are classification problems and a real dependence of the
B/T ratio on galaxy luminosity that contributes to the
scatter. Therefore also the B/T ratio is not a classification
criterion equivalent to Hubble type, and it is not equivalent either
to the concentration index criterion, as shown in
Fig.~\ref{fig:bt_c31}, where we see the presence of a rather large
scatter in the value of B/T for any given value of C$_{31}$.

Keeping in mind these differences, and the fact that all three form
criteria (Hubble type, concentration index, and B/T ratio) share some
common basis in terms of their ability to differentiate among galaxies
with different appearance, structure, and stellar populations, we
choose to base our ``form'' classification on the concentration index
C$_{31}$. With respect to Hubble type it has the advantage of being a
quantitative index, that can be easily measured for all galaxies
independently from the observer's viewing angle, and is less affected
by distance related resolution problems. With respect to the B/T
ratio, it has the advantage of being model independent, not requiring
the decomposition of a galaxy into a bulge and a disk component, a
process which is always affected by considerable uncertainties.
Another advantage offered by the concentration index is its continuous
distribution of values, as compared to the quantized distribution of
Hubble types, and the peculiar distribution of B/T values (composite
B+D photometric profiles show a continuous distribution of values, but
pure de Vaucouleurs profiles have B/T equal to 1, and pure exponential
disk profiles have B/T equal to 0 by definition).

Finally we remark that, at least in the $H$-band, where internal dust
extinction effects are quite negligible, galaxy viewing angle has
little influence on the derivation of C$_{31}$ values. The
distribution of C$_{31}$ covers the full range of observed values at
all galaxy inclinations (as derived from the observed axial ratio),
but we observe a moderately significant trend toward higher C$_{31}$
values at high inclination. The median C$_{31}$ value for galaxies
with inclination below 50 degrees is 2.99, while for galaxies with
higher inclination is 3.56 (in both cases the distribution has a
standard deviation of approximately 1.5).


\begin{figure*}
\resizebox{\hsize}{!}{\includegraphics{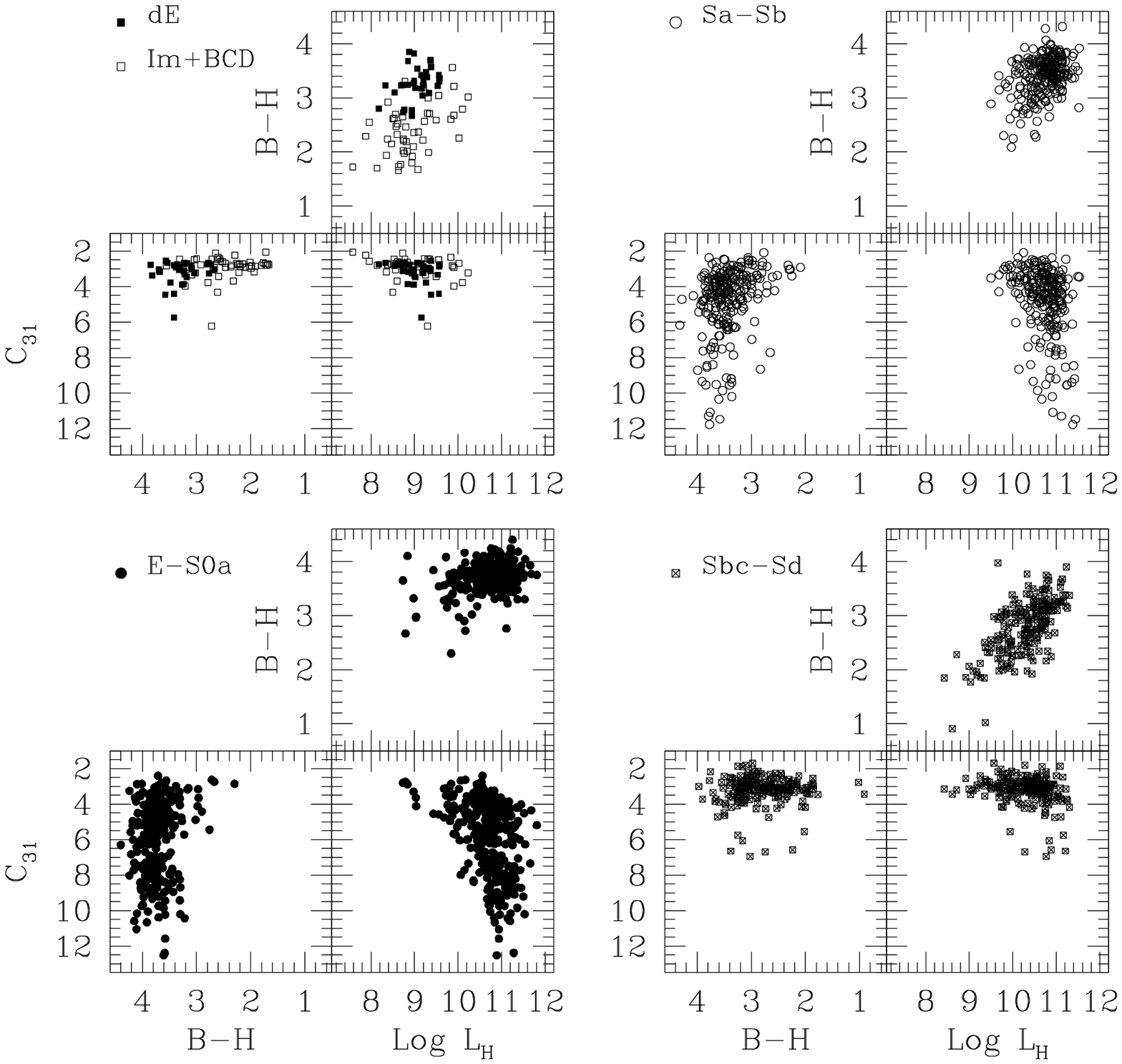}}
\caption{The distribution of $H$-band luminosity $L_H$, $B-H$ color,
and concentration index C$_{31}$ for the galaxies in our sample,
divided into 4 groups of morphological types. In the left half of the
plot we present dwarf galaxies (both dE and dIrr) on the top part and
E plus S0 galaxies on the bottom part. In the right half of the plot
we present Sa and Sb galaxies on the top part, and Sc and Sd galaxies
on the bottom part.}
\label{fig:lcube_e_sc}
\end{figure*}

\begin{figure*}
\resizebox{\hsize}{!}{\includegraphics{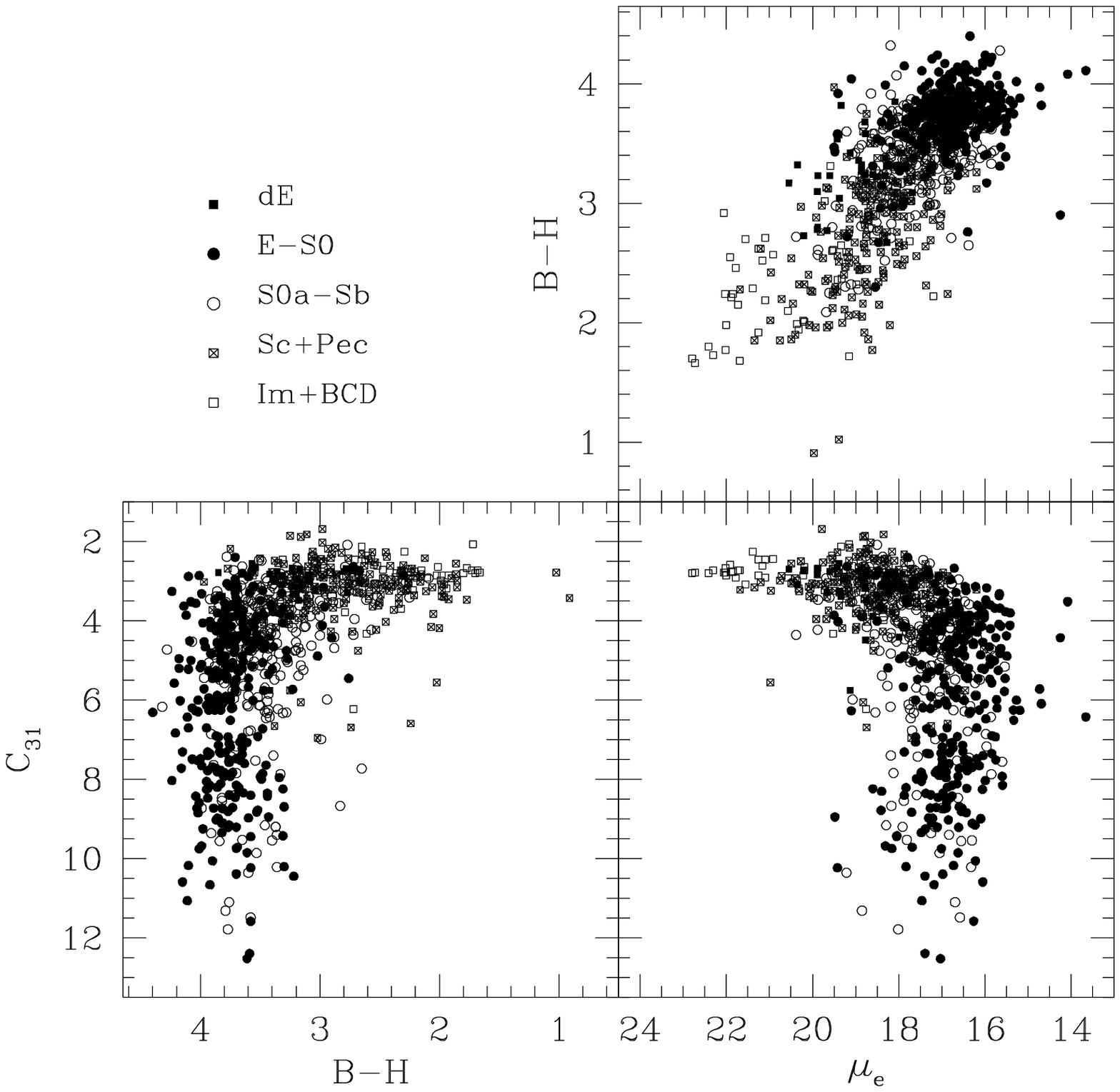}}
\caption{The distribution of effective surface brightness
$\mu_\mathrm{e}$, $B-V$ color, and concentration index C$_{31}$ for
the galaxies in our sample. As in the previous figure, different
symbols identify galaxies within different ranges of Hubble types.}
\label{fig:mucube}
\end{figure*}

\section{Scale, form, and color}
\label{sec:cube}

Whitmore (\cite{Whit84}) introduced the first quantitative galaxy
classification scheme that incorporates both structural and spectral
parameters, introducing the concepts of galaxy ``scale'' (in his
analysis a combination of absolute blue luminosity and isophotal
radius) and ``form'' (in his analysis a combination of $B-H$ color and
B/T light ratio). He concluded that these two parameters would provide
an optimal classification scheme for spiral galaxies, to which his
work was limited. His choice for the specific combinations that define
the form and scale parameters was based on the observation that those
quantities show the strongest correlations between themselves among
the 30 parameters taken into consideration in his work. More recently
BJC have presented a quantitative classification scheme along the same
lines, extended to galaxies of all morphological types. However they
need three parameters to obtain a satisfactory classification of all
galaxies: a spectral index ($B-V$ color), a scale (surface
brightness), and a form (concentration index or image asymmetry)
parameter.

\subsection{The luminosity, color, C$_{31}$ cube}

Independently from BJC, we have defined a very similar framework to
describe the main structural and photometric properties of normal
galaxies. Our sample in fact does not include a significant fraction
of starbursting, interacting, or extremely low surface brightness
objects. In agreement with BJC, we conclude that at least three
parameters are needed to provide a unique classification for all
galaxies. Figs.~\ref{fig:lcube} and \ref{fig:mucube} show two
alternatives definitions for such a classification cube. The first
one, presented in Fig.~\ref{fig:lcube} is based on $H$-band
luminosity, color (either $B-H$ or $B-V$, this second not shown in the
figure), and concentration index C$_{31}$. In the second one,
presented in Fig.~\ref{fig:mucube}, the effective surface brightness
$\mu_\mathrm{e}$ replaces the $H$-band luminosity. To keep the
connection with the Hubble types classification open, in both figures
different symbols identify galaxies within different ranges of Hubble
types.

It is well known that both early- and late-type galaxies obey a
color-luminosity relation (see, for example, Visvanathan \& Sandage
\cite{VS77}; Bower et al. \cite{BLE92} for early-type galaxies, and
Tully et al. \cite{TMA82}; Gavazzi et al. \cite{Gav96c} for spiral
galaxies).  Since the two relations are quite different, their origins
are also believed to be different. The small color changes of
early-type galaxies are most likely produced by metallicity variations
within a coeval galaxy population (Arimoto \& Yoshii \cite{AY87}; see
however Worthey et al. \cite{Wor96} for a discussion on the effects of
age variations within a metal homogeneous population).  The much
broader color range covered by late-type galaxies is mainly the result
of a broad range of star formation histories (Searle et
al. \cite{SSB73}; Kennicutt \cite{Kenn83}; Sandage \cite{San86}),
coupled with an equally broad range of dust extinction values. This
color-luminosity correlation is well visible in the upper panel of
Fig.~\ref{fig:lcube}.

\begin{figure*}
\resizebox{\hsize}{!}{\includegraphics{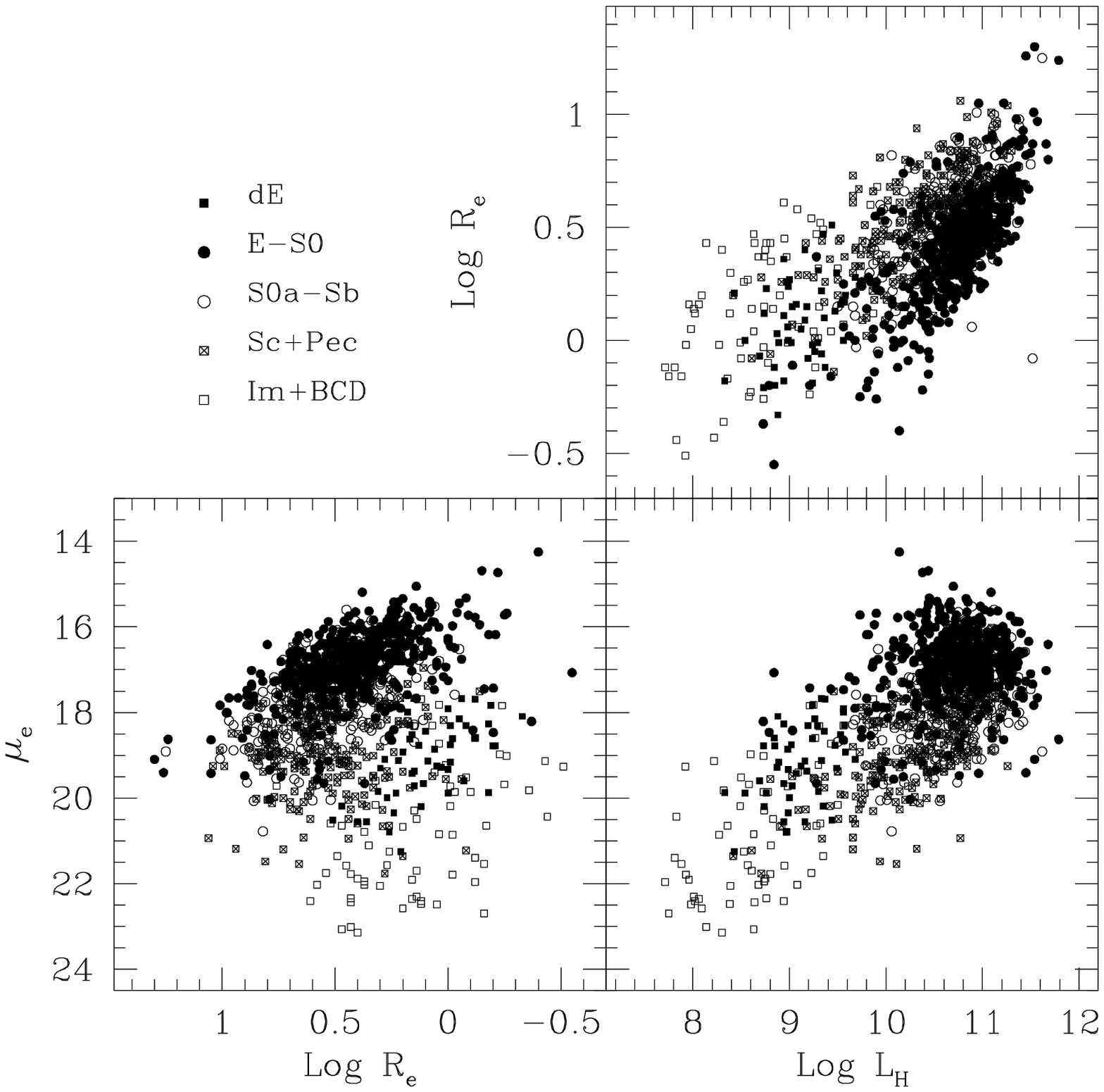}}
\caption{The distribution of $H$-band luminosity $L_H$, effective
surface brightness $\mu_\mathrm{e}$, and effective radius
$r_\mathrm{e}$ for the galaxies in our sample. Different symbols
identify galaxies within different ranges of Hubble types, as in the
previous figures.}
\label{fig:pfp}
\end{figure*}

BJC have also already described a correlation between color and
$B$-band concentration index, although they did not discuss what we
consider to be the main feature visible in the lower panels of
Figs.~\ref{fig:lcube} and \ref{fig:mucube}, which is the strong
non-linearity in the correlation between C$_{31}$ and either color,
$H$-band luminosity, or effective surface brightness. It is clearly
evident from these figures that there are two main regimes of galaxy
properties, that produce two almost orthogonal distributions in the
parameter space we are considering. At low luminosity
($L_\mathrm{H}<10^{10}L_{\odot}$) all galaxies consistently have very
small C$_{31}$, independently from their color (and from their Hubble
type, for that matter), while only at high luminosity
($L_\mathrm{H}>10^{10}L_{\odot}$) we observe galaxies with large
values of C$_{31}$. At the same time blue galaxies consistently have
low values of C$_{31}$, and it is only among the red ones that large
values of C$_{31}$ are present. Alternatively, we can say that
galaxies with small C$_{31}$ can have all colors and all luminosities,
while those with a large C$_{31}$ are only found among red luminous
objects. We stress here the fact that this qualitative picture is not
sensitive to the choice of photometric bands used to derive the color
parameter, at least as far as optical or optical-near infrared colors
are involved.  The finding of a luminosity threshold for the
appearance of high C$_{31}$ values is in agreement with the finding
discussed in Papers V and VII, that pure de Vaucouleurs profiles are
absent from the classification of photometric profiles below the
luminosity $L_\mathrm{H} \simeq 10^{10}L_{\odot}$.

\begin{figure*}
\resizebox{\hsize}{!}{\includegraphics{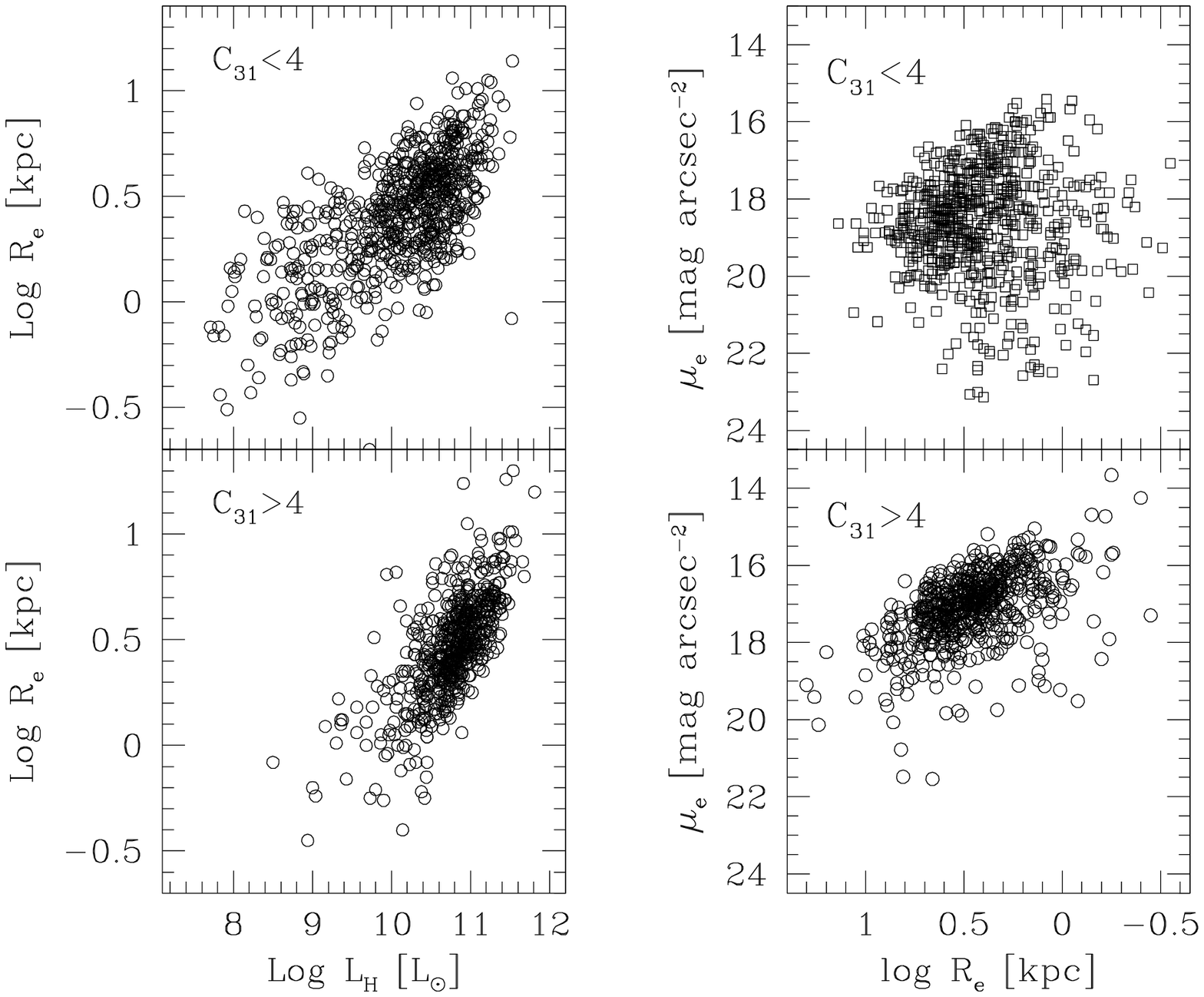}}
\caption{The correlation between $H$-band luminosity $L_H$ and
effective radius $r_\mathrm{e}$ (left panels), and that between
effective radius $r_\mathrm{e}$ and effective surface brightness
$\mu_\mathrm{e}$ (right panels) for the galaxies in our sample,
divided into two groups according to their C$_{31}$ values. The top
panels shows galaxies with C$_{31}\leq$4, while the bottom one shows
those with C$_{31}>$4.}
\label{fig:rlum_korm_c31}
\end{figure*}

As discussed in Sect.~\ref{sec:data} our sample is composed mainly
of galaxies from the Coma Supercluster and the Virgo Cluster. The
different distance of these two structures translates into different
completeness limits when luminosity is taken into consideration. To
provide a more reliable indication of the relative density of galaxies
in the various parts of our classification cube we reproduce it again
in Fig.~\ref{fig:lcube_virgo}, plotting in this case only Virgo
Cluster galaxies. Notice that even in this case, however, for
luminosities $L_\mathrm{H} < 10^{9}L_{\odot}$ sample incompleteness
becomes non-negligible, and the density of galaxies in the plot does
not represent any more the true fraction of objects that lay in this
part of the galaxies' parameter space. The most striking difference
between Figs.~\ref{fig:lcube} and \ref{fig:lcube_virgo} concerns spiral
galaxies. Within the global sample the color-magnitude relation for
these objects appears much steeper than in the more complete Virgo
sample, purely as a result of sample incompleteness. In fact our
sample is an optically selected one, complete down to a certain
optical magnitude (as discussed in Sect.~\ref{sec:data}). Therefore
when one considers $H$-band data, the magnitude limit for the sample
changes with galaxy color, shifting to brighter luminosities as the
color changes from blue to redd (a diagonal line in
Fig.~\ref{fig:lcube}).

To better illustrate the relation between the classification cube
presented here and the classic Hubble classification scheme, we present
in Fig.~\ref{fig:lcube_e_sc} the same plots as in Fig.~\ref{fig:lcube},
separating the galaxies in different panels according to their Hubble
type. It is evident from these figures that the scatter within the
C$_{31}$-Hubble type relation can be partly explained by the luminosity
(color) dependence of the concentration index, that is quite evident
even when a restricted range of Hubble types is considered.

\subsection{The $\mu_\mathrm{e}$, color, C$_{31}$ cube}

An almost equivalent picture to that presented in Fig.~\ref{fig:lcube}
is obtained substituting the effective surface brightness
$\mu_\mathrm{e}$ to the $H$-band luminosity (see
Fig.~\ref{fig:mucube}). There are nonetheless two noticeable
differences between these two data cubes. The first one is the smaller
scatter shown by the $\mu_\mathrm{e}$-color relation with respect to
the luminosity-color one. This results mainly from a lack of low
$\mu_\mathrm{e}$ ellipticals, while late-type galaxies show similar
behaviour in the two diagrams.  This effect could however be due to an
observational bias, as red objects with $\mu_\mathrm{e}$ fainter than
20 H--mag~arcsec$^{-2}$ are very difficult to observe and detect in
the near infrared. As a result our sample is strongly incomplete with
respect to dwarf-elliptical galaxies.  Should the smaller scatter be
confirmed even after the completion of our survey, it could probably
be explained by the fact that $\mu_\mathrm{e}$ is not purely a scale
parameter like luminosity, but a hybrid one, mixing both scale and
form in its definition.  The second difference is the total absence of
a correlation between $\mu_\mathrm{e}$ and C$_{31}$ for high C$_{31}$
objects, compared to a small but significant trend between luminosity
and C$_{31}$ (the mean $log L_H$ for objects with $4<$C$_{31}<5$ is
10.6, while for objects with C$_{31}>9$ it is 11.0). Another
interesting result is the fact that the observed strong correlation
between $\mu_\mathrm{e}$ and color, which confirms earlier results
presented by Gavazzi et al. (\cite{Gav96c}), is in disagreement with
the results presented by BJC (see the upper panels of their Fig. 3),
that confirm instead the so-called Freeman's law (Freeman
\cite{Free70}). This is yet another indication that the validity of
Freeman's law is limited to one specific photometric band.

It could be argued that this second classification cube is better
suited than the one discussed in the previous section to represent
normal galaxy properties, both because of the smaller scatter observed
in the $\mu_\mathrm{e}$-color relation with respect to the $L_H$-color
one (should this feature be confirmed when observations of our sample
will be completed), and because surface brightness is a
distance-independent quantity (cosmological dimming correction aside),
whereas luminosity is not.  However what could be better in terms of a
classification scheme might not be equally relevant in building a
physical model for galaxy formation and evolution. In that case total
near infrared luminosity, which is a good tracer of galaxy mass,
should be certainly considered as a more fundamental parameter than
surface brightness.

\begin{figure*}
\resizebox{\hsize}{!}{\includegraphics{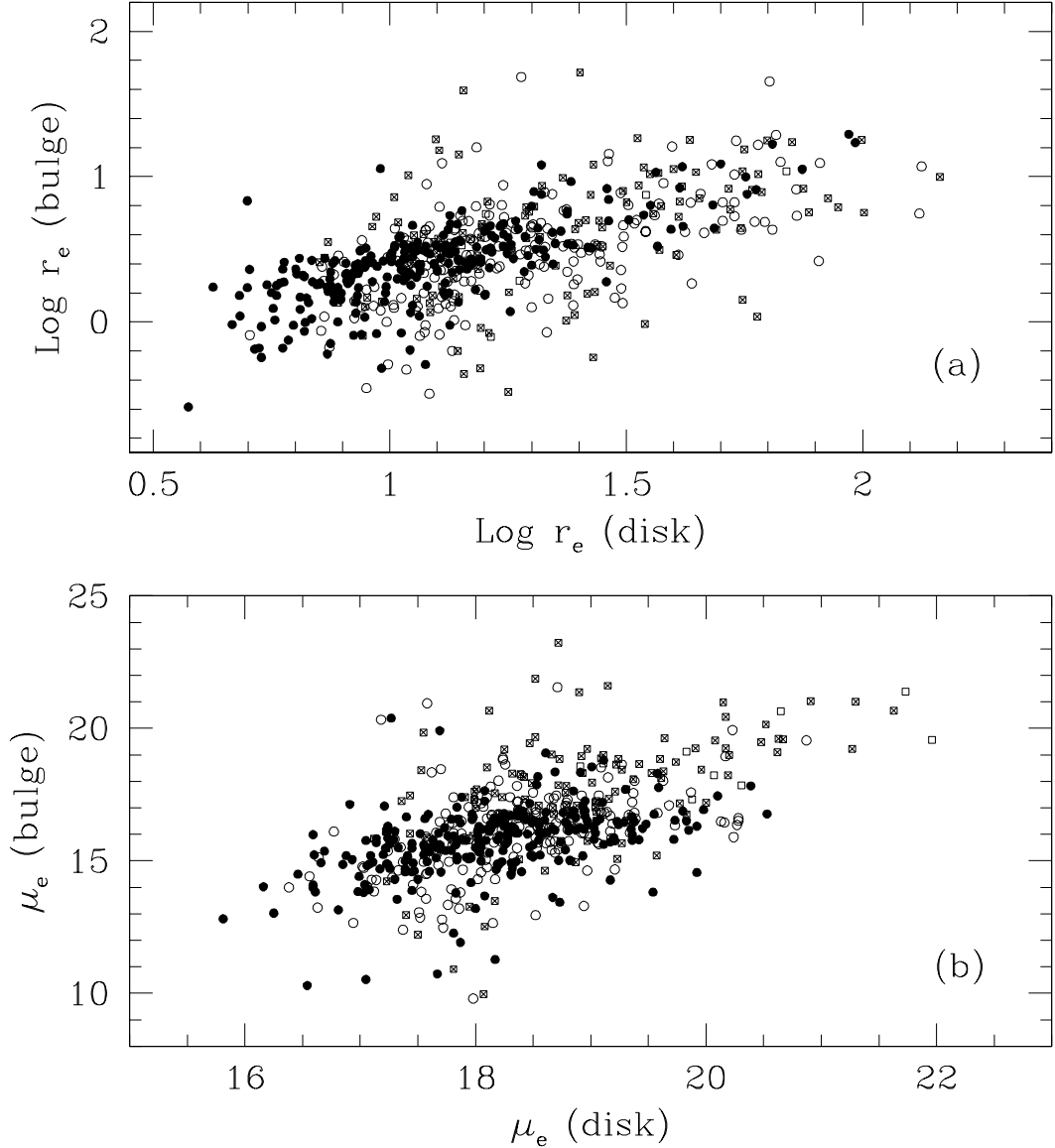}}
\caption{The correlation between bulge and disk effective radius (a)
and between bulge and disk effective surface brightness (b). Plotting
symbols are as in Fig.~\ref{fig:lcube}: filled circles represent E
and S0 galaxies, empty circles Sa to Sb galaxies, crossed squares Sc
to Sdm galaxies, open squares Im and BCD galaxies.}
\label{fig:bulge_disk}
\end{figure*}

\begin{figure*}
\resizebox{\hsize}{!}{\includegraphics{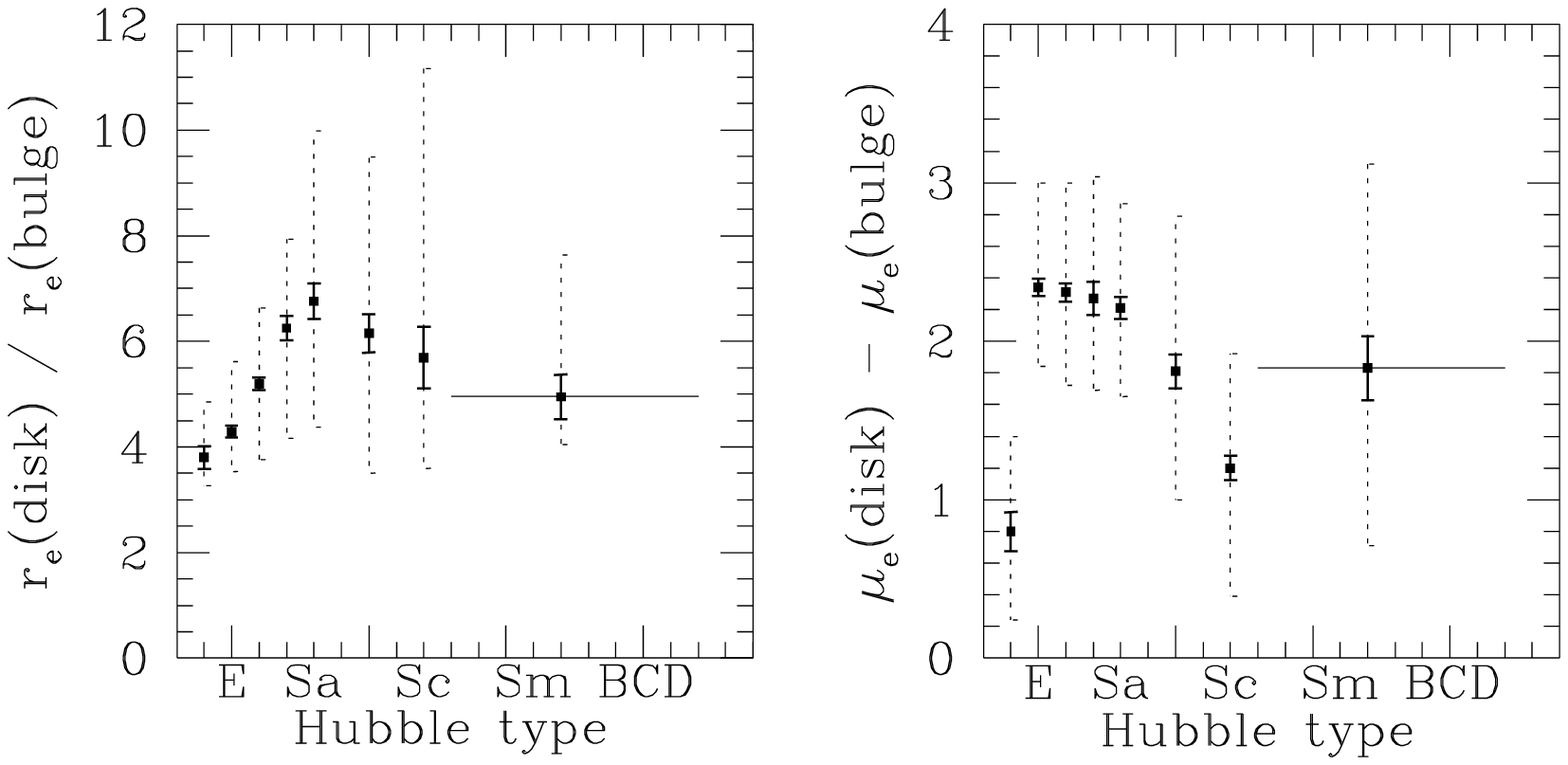}}
\caption{The median of the ratio between disk and bulge effective
radius (a) and of the difference between disk and bulge effective
surface brightness (b) for different morphological types (filled
squares), with the associated statistical uncertainty and the
inter-quartile range of their value distribution (dased vertical
lines). Types later than Sc are grouped together to improve
statistics.}
\label{fig:bd_type}
\end{figure*}

\section{The Photometric Fundamental Plane}

The possibility of replacing $H$-band luminosity with effective
surface brightness in the classification cube, as discussed above, is
due to the presence of a significant correlation between those two
quantities. In fact a number of correlations are known to exist, at
least for early-type galaxies, that involve luminosity, effective
surface brightness, and effective radius.  Luminosity correlates with
effective radius (Fish \cite{Fish64}), stellar velocity dispersion
(Faber \& Jackson \cite{FJ76}), and effective surface brightness
(Binggeli et al. \cite{Bing84}), while effective radius and effective
surface brightness correlate among themselves (Kormendy \cite{Kor77};
see Guzm{\'a}n et al. \cite{Guz93} for a global discussion about these
correlations).  These relations exhibit a larger scatter than can be
accounted for by measurement errors alone, and this lead to the idea
that early-type galaxies populate a plane in the 3--parameter space of
effective radius, effective surface brightness and stellar velocity
dispersion, independently introduced by Djorgovski \& Davis
(\cite{DD87}) and by Dressler et al. (\cite{7S}). Modified versions of
this Fundamental Plane relation have since then been proposed, that
use color (de Carvalho \& Djorgovski \cite{deCar89}) and luminosity
(Scodeggio et al. \cite{SGH97}) as a substitute for the stellar
velocity dispersion.

A Photometric Fundamental Plane correlating galaxy luminosity,
effective radius $r_\mathrm{e}$ and effective surface brightness
$\mu_\mathrm{e}$ still exists\footnote{although one must keep in mind
that only two of the three parameters involved are statistically
independent, as a result of our definition of effective radius and
effective surface brightness: given the total flux $f_\mathrm{t}$ and
the effective radius $r_\mathrm{e} = r(f_\mathrm{t}/2)$, we have
$\mu_\mathrm{e}= -2.5\log(f_\mathrm{t}/2\pi r_\mathrm{e}^2) +
Z_\mathrm{p}$} even when considering galaxies of all morphological
types, as shown in Fig.~\ref{fig:pfp}. However, as it could be
expected, the picture one obtains when considering galaxies of all
types is more complex than the one provided by early-type objects
alone. While the shape and scatter of the luminosity-$\mu_\mathrm{e}$
relation are not influenced by the sample morphological mix, both the
$r_\mathrm{e}$-luminosity and the $r_\mathrm{e}$-$\mu_\mathrm{e}$
relation show a significant segregation of galaxies as a function of
their concentration index (or, alternatively, their morphological
type).  In the case of the $r_\mathrm{e}$-luminosity relation the
difference in galaxy properties between low- and high-C$_{31}$
galaxies is not extreme, and both classes of objects follow a well
defined correlation, although with significantly different slopes and
scatter (see the left panels in Fig.~\ref{fig:rlum_korm_c31}). The
effect is more evident in the case of the \re-$\mu_\mathrm{e}$
correlation, as shown in the right panels of
Fig.~\ref{fig:rlum_korm_c31}. It is clear from the figure that only
for high C$_{31}$ galaxies, irrespective of their morphological type,
$r_\mathrm{e}$ and $\mu_\mathrm{e}$ are correlated to produce the well
known Kormendy relation of early-type galaxies. For low C$_{31}$
galaxies, instead, no correlation is observed between the two
parameters, except for a lack of objects with large $r_\mathrm{e}$ and
very bright $\mu_\mathrm{e}$ (the bottom left corner of the diagram).
This result is in qualitative agreement with the finding of Graham \&
de Block (\cite{GdeB01}) that very late-type spirals (thus galaxies
with low C$_{31}$) have smaller and fainter disks than earlier-type
ones. The zone of avoidance at large $r_\mathrm{e}$ and very bright
$\mu_\mathrm{e}$, which is common to galaxies of all C$_{31}$ values,
has been already discussed (for the case of disk galaxies only) by
Dalcanton et al. (\cite{Dal97}; see their Fig. 4), and has its origin
in the existence of an exponential cutoff at the high-mass end of the
mass function of galaxies (closely mirrored by the cutoff in the
galaxy luminosity function).

\section{Bulge and disk parameters}
\label{sec:bulge_disk}

The main motivation for using a concentration index in our
classification scheme instead of something like a bulge to disk light
ratio was the fact that measurements of C$_{31}$ do not require
model-dependent decompositions of a galaxy light distribution, which
carry with them systematic uncertainties very difficult to
estimate. Still those decompositions are routinely carried out, and
are used to analyze in fine detail models of galaxy
formation. Courteau et al. (\cite{Court96}) were the first to report
the existence of a correlation between disk and bulge scale lengths
for two samples of late-type spirals, that they interpreted as
evidence for the presence of secular evolution in driving bulge
formation. Their result has been recently confirmed also by
Khosroshahi et al. (\cite{Khos00}). However these authors, as well as
Graham \& Prieto (\cite{GP99}), find systematic changes in the ratio
of disk to bulge scale length as a function of morphological type, and
they interpret this result as evidence against the formation of bulges
via secular evolution of the disk of spiral galaxies.

We have analyzed the relation between bulge and disk parameters for
the 667 galaxies in our sample that required a bulge+disk
decomposition of their surface brightness profile. We find that both
the disk and bulge effective radius and disk and bulge effective
surface brightness are correlated, although with a large scatter.
These correlations are presented in Fig.~\ref{fig:bulge_disk}, where
different symbols identify galaxies of different morphological type
(symbols are the same as in Figs.~\ref{fig:lcube} and
\ref{fig:mucube}).  There is also some marginal evidence for
systematic changes with morphological type in the ratio between the
two radii, or the difference between the two surface brightnesses, as
shown in Fig.~\ref{fig:bd_type}. In contrast with the finding of
Graham \& Prieto (\cite{GP99}), we find that Sa galaxies have a larger
ratio between disk and bulge effective radius than later-type spirals,
and they also have a much larger difference in effective surface
brightness than later types. This difference cannot be considered
really significant, as Graham \& Prieto (\cite{GP99}) themselves have
shown that a change in the function used to fit the surface brightness
profile of the bulge (an exponential profile instead of a S{\'e}rsic
one) leads to a complete reversal of their results, bringing them in
agreement with ours. This same caveat applies to the comparison with
the results obtained by Courteau et al. (\cite{Court96}): they used a
fixed profile shape for fitting the bulge photometric profile, while
we have chosen either a de Vaucouleurs $r^{1/4}$ or an exponential
disk profile according to an objective $\chi^2$ criterion. The large
scatter which is superimposed on the correlations shown in
Fig.~\ref{fig:bulge_disk} seems to argue against the presence of a
single, homogeneous mechanism that could regulate the bulge formation
and its interaction with the surrounding disk.

\section{Discussion and conclusions}
\label{sec:conclusions}

A classification scheme for any kind of object, either real or
totally abstract ones, is really useful only if it can be used to sort
properties of those objects other than those explicitly used for the
classification itself. For example, the Harvard stellar spectra
classification scheme, created purely on the basis of spectral shape
and spectral line patterns, is still in use because it was later
recognized that it sorts stars in terms of surface temperature, and
ultimately in terms of mass. As far as galaxies are concerned,
Hubble's tuning fork scheme is by far the most commonly use
classification tool. The origin of its success, besides the fact that
it is relatively easy to use, is mainly due to the fact that it helps
sort important physical galaxy properties, like current stellar
population and star formation history, or gas and dust content.

However this sorting is only a partial one, and while clear trends
exist from many galaxy properties as a function of Hubble's type, it
is also true that these properties show a large scatter within each
type, so that there is a large overlap in their distribution between
different types (see Roberts \& Haynes \cite{RH94} for a recent
review). We have shown two examples of this situation in
Fig.~\ref{fig:c31_bt_type}. These large scatters naturally lead to the
idea of including some new parameters in the classification scheme, to
increase its discriminating power. However, other limitations within
Hubble's scheme, namely its qualitative and quantized nature, and its
failure to classify a significant fraction of known galaxies, severely
limit the usefulness of any such extension. We have therefore decided
to explore the possibility of replacing Hubble's scheme
altogether. This is not a new idea, as complete replacements were
already proposed in the past, most noticeably by Morgan(\cite{Mor58},
\cite{Mor59}), Whitmore (\cite{Whit84}), and most recently by
BJC. Building on Whitmore's ideas, we have independently obtained a
classification scheme that turns out to be qualitatively very similar
to the one proposed by BJC, and is illustrated by
Fig.~\ref{fig:lcube}.

Contrary to BJC, we do not make here any attempt at reproducing the
sub-division among Hubble's types within our scheme, as we are only
interested in building a scheme that can give a complete description
of galaxy properties based on a small number of quantitative and 
easy-to-measure parameters. In agreement with BJC we find that normal
galaxies (we do not have peculiar, strongly interactive, or extremely
low surface brightness galaxies in our sample) can be described by a
three parameter model. Using Whitmore terminology, we can call these
three parameters galaxy scale (for us the $H$-band luminosity), form
(for us the concentration index C$_{31}$), and color (we use the $B-H$
color, but any optical or optical-near infrared color can be used for
this purpose). The need for three parameters is clearly shown by
objects that belong to the two different property regimes existing
along the strongly non-linear relations between C$_{31}$ and either color
or luminosity. At low luminosity (see the bottom panels of
Fig.~\ref{fig:lcube}) we need color information to distinguish
between a dE and a dIrr galaxy with the same C$_{31}$, while at high
luminosity we need C$_{31}$ information to distinguish between an S0 and
an Sc galaxy with the same color.

The three parameters are relatively easy to derive from photometric
observation data, and do not require any modeling of the galaxy
light distribution. However the measurement of C$_{31}$ imposes
constraints on the resolution of the imaging data that could be used
to classify galaxies, and it is clear that even using our model it
would be extremely difficult to rely on ground-based observations to
classify high-redshift galaxies. 

An alternative representation of the classification cube can be
obtained substituting galaxy luminosity with effective surface
brightness, taking advantage of the correlation existing between the
two quantities (see Figs.~\ref{fig:mucube} and \ref{fig:pfp}). This is
perhaps the most unexpected result of our analysis, as it is in
contrast with previous results based on observations carried out at
shorter wavelengths. We find that a single, well-defined relation
exists between luminosity and $\mu_\mathrm{e}$ in the $H$-band, as
opposed to the two separate regimes observed by Binggeli et
al. (\cite{Bing84}) in the $B$-band, and between $B-H$ or $B-V$ color
and $\mu_\mathrm{e}$, as opposed to the much shallower relation
between $B-V$ and $\mu_\mathrm{e}$ in the $B$-band presented by
BJC. Actually, for $B-V < 0.8$ the BJC relation is completely flat, as
demonstrated by the vertical boundary between late- and
intermediate-type galaxies in the upper panel of their Fig. 3, and as
should be expected on the basis of Freeman's law (Freeman
\cite{Free70}).  On the contrary, we find that surface brightness in
the near-infrared is not a scale-free parameter, but it depends on the
galaxy luminosity, confirming earlier results by Gavazzi et
al. (\cite{Gav96c}) and de Jong (\cite{deJ96}).

These differences highlight the importance of selecting the ``right''
photometric band for mapping the structural properties of galaxies. In
this respect near-infrared bands have two key advantages over optical
ones: reduced dust extinction effects, and reduced effects of current
star formation activity on both the galaxy total luminosity and
surface brightness distribution. For these reasons, near-infrared
luminosity traces quite accurately galaxy mass (see for example
Gavazzi et al. \cite{Gav96c}, and Scodeggio et al. \cite{tilt}). Using
this luminosity-mass equivalence we can obtain an accurate picture of
the role that galaxy mass has in determining the global structure of a
galaxy, via the influence it has in determining the galaxy luminous
matter mean density and global distribution.

If the observed correlation between mass and surface brightness is
already an indication in favor of a scale-dependent galaxy collapse
mechanism, the strong non-linearity of the relation between mass and
concentration index C$_{31}$ should pose very strong constraints on
such a scenario. Strong light concentrations (high C$_{31}$ values),
that are somewhat correlated to the presence of a dominant spheroidal,
bulge-like component (but not in a simple way, as the scatter in the
C$_{31}$-B/T ratio seen in Fig.~\ref{fig:bt_c31} demonstrates), appear
only for luminous red galaxies, at $L_H > 10^{10}~L_{\odot}$. Below
that limit galaxies do not have enough mass to drive the concentration
of mass and luminous matter towards their center. Above the limit this
concentration is possible, but it does not take place automatically
for all galaxies. If it happens, however, it must happen quite early
during the evolution of a galaxy, or at least involve only old stellar
populations without the onset of new star formation activity. In fact
high values of C$_{31}$ are present only among very red galaxies that
must have formed a large fraction of their stars during the first few
(one to three) billion years of their history.

This important role of galaxy mass in determining the galaxy structure
matches a similar role that mass has in determining the star formation
history of galaxies, at least for spiral ones (see Gavazzi \&
Scodeggio \cite{GS96} and also Boissier \& Prantzos
\cite{BP00}). Unfortunately the link between galaxy mass and
near-infrared luminosity is the only well established one along the
path from a phenomenological classification scheme to a self
consistent physical model, and we don't have other simple physical
properties that can be unequivocally associated with star formation
(\ie color) and matter distribution within a galaxy, although it is
obvious that angular momentum must be playing an important role in
determining galaxy structure (see, for example, Dalcanton et
al. \cite{Dal97}; Boissier \& Prantzos \cite{BP00}; Pierini et
al. \cite{Dan01}). It is therefore not possible to build, based on the
present data, a simple physical model that could be used, for example,
to explore the origin of the scatter that is still present in all
relations involving galaxy structural parameters.


\end{document}